\def\rfr#1{(\ref{#1})}

\def\bar{\begin{eqnarray}}
\def\ear{\end{eqnarray}}

\def\eqi{\begin{equation}}
\def\eqf{\end{equation}}
\def\eqia{\begin{eqnarray}}
\def\eqfa{\end{eqnarray}}
\def\rp#1#2{{#1\over#2}}
\def\ct#1{\cite{#1}}
\def\lb#1{\label{#1}}

\def\oc2{$\mathcal{O}(c^{-2})$}

\def\bds#1{\boldsymbol{#1}}

\documentclass[11pt]{article}
\usepackage{amsmath,amsthm,amscd,amssymb}
\usepackage{latexsym}
\usepackage{graphicx,epsfig}

\begin{document}

\noindent{\bf \LARGE{On the impossibility of measuring the general
relativistic part of the terrestrial acceleration of gravity with
superconducting gravimeters }}
\\
\\
\\
{Lorenzo Iorio}\\
{\it Viale Unit${\grave{a}}$ di Italia 68, 70125\\Bari, Italy
\\e-mail: lorenzo.iorio@libero.it}

\begin{abstract}
In this paper we very preliminarily investigate the possibility of
measuring the post-Newtonian general relativistic gravitoelectric
and gravitomagnetic components of the acceleration of gravity on
the Earth, in continuous regime, with
%
%
two absolute measurements
%
%
at the equator and the south pole with superconducting
gravimeters. The magnitudes of such relativistic effects are
10$^{-10}$ m s$^{-2}$ and $10^{-11}$ m s$^{-2}$, respectively.
Unfortunately, the present-day uncertainties in the Earth's
geodetic parameters which enter the classical Newtonian terms
induce systematic errors 1-2 orders of magnitude larger than the
relativistic ones. Moreover, a $\sim 1 $ ngal sensitivity can be
reached by the currently available superconducting gravimeters,
but only for relative measurements.
\end{abstract}

\section{Introduction}
%

The classical test-bed of the Einstein's General Theory of
Relativity (GTR), in its linearized weak-field and slow-motion
approximation \ct{soffel} valid in the Earth's neighborhood and
throughout the Solar System, has always been represented by the
motion of test particles and electromagnetic waves in the
gravitational field of massive astronomical bodies \ct{will}. This
is so due to the extreme smallness of the relativistic effects
with respect to the much larger Newtonian features of gravity.

However, in the more or less recent past various Earth-based
laboratory experiments \ct{bra, pippard, bra2, vitale, tartrug02,
stedman, ioriopolo} have been proposed, mainly to detect the
elusive gravitomagnetic part of the gravitational field of the
Earth induced by its proper angular momentum (see Section
\ref{gtr}). Up to now, they have not been performed, mainly due to
the extreme difficulty both to reach the necessary high
sensitivity and to cope with the numerous and much larger
classical noising effects of natural and cultural origin.

In regard to the possibility of testing GTR with terrestrial
geodetic techniques \ct{soffel}, gravimetry can allow to penetrate
into the relativistic regime. Among the most sensitive apparatus
there are the superconducting gravimeters (SG) \ct{goodk}. Their
most remarkable quality is their high immunity to the
environmental temperature variations, that generally determine
variations of the elastic constant of the mechanical spring which,
indeed, is replaced by a magnetic levitation of a superconducting
sphere in the magnetic field of superconducting, persistent
current coils. The goal is to utilize the high stability of such
currents to create a perfectly stable spring. The magnetic
levitation is designed to provide independent adjustment of the
total levitating force and the force gradient so that it can
support the full weight of the sphere and still yield a large
displacement for a small change in gravity. Recent developments in
such a field have pushed the accuracy of SG down to $10^{-12}
g=10^{-11}$ m s$^{-2}=$ 1 ngal \ct{nanogal} for relative
measurements. For some SG performed tests of certain
non-relativistic preferred-frame and preferred-locations effects,
in the context of the Parameterized Post Newtonian (PPN)
formalism, see Section 8.2 of \ct{WIL}. However, it must also be
noted that there are great difficulties to use SG in absolute
measurements due to calibration issues.

\section{The post-Newtonian general relativistic components of the acceleration of
gravity}\label{gtr} The acceleration experienced by a test mass in
the gravitational field of a central body, to the post-Newtonian
\oc2\ level of GTR in its linearized weak-field and slow-motion
approximation, is represented by a gravitoelectric component, due
to the Schwarzschild part \ct{Schw} of the spacetime metric, and a
gravitomagnetic component, due to the Lense-Thirring part
\ct{leti} of the spacetime metric. We can write it as $\bds
a\equiv\bds a_{\rm Newton }+\bds a_{\rm Schwarzschild}+\bds a_{\rm
Lense-Thirring}$ with \ct{soffel}
\eqi \left\{
\begin{array}{lll}\lb{accel}
\bds a_{\rm Newton }&=&-\rp{GM}{r^3}\bds r,\\\\\bds a_{\rm
Schwarzschild}&=&
\rp{GM}{c^2r^3}\left\{\left[\rp{4GM}{r}-v^2\right]\bds r+4(\bds
r\cdot\bds v )\bds v\right\},\\\\
\bds a_{\rm Lense-Thirring}&=&\rp{2G}{c^2 r^5}\left[3(\bds
r\cdot\bds J)\bds r\times\bds v+r^2\bds v\times\bds J\right].
\end{array}
\right. \eqf
$G$ is the Newtonian gravitational constant, $c$ is the speed of
light in vacuum, $M$ is the mass of the central body, $\bds J$ is
its proper angular momentum\footnote{For a homogeneous spherical
body of mass $M$, radius $R$ spinning at a rate $\omega$
$J=(2/5)MR^2\omega$; for the Earth the factor $2/5=0.4$ is
replaced by $0.330841$ \cite{iers}.}, $\bds r$ and $\bds v$ are
the position and velocity vectors, respectively, of the test mass.
It is assumed that the origin of the adopted inertial reference
frame is located at the centre of mass of the body of mass $M$.
\section{Absolute measurements at the pole and the
equator} We will now explore the possibility of measuring the
gravitoelectric and gravitomagnetic components of the terrestrial
gravitational field by measuring the acceleration of gravity at
the equator and at the South Pole and making a cross-check between
these two absolute measurements. To this aim, let us, now,
consider a test body fixed on the Earth'surface, so that $\bds
v=\bds\omega\times R\bds{\hat{r}}$ and $v=\omega R\cos\lambda$
where $\bds \omega=\omega\bds k$ is the Earth's angular velocity
vector, $\bds{\hat{r}}$ is a unit vector from the Earth's centre
to the location of the test mass, $\lambda$ is the geocentric
latitude and $R$ is the Earth's radius at the location of the test
body. From \rfr{accel} it is easy to see that at the equator
$(\bds r\cdot\bds J=0)$ the Lense-Thirring acceleration is
entirely radial and directed outward, as the centrifugal
acceleration, while at the South Pole it is absent ($\bds v=0$).
The gravitoelectric Schwarzschild component is always directed
radially because at the equator $\bds r\cdot \bds v=0 $ and at the
poles $\bds v=0$. Then, the acceleration of gravity can be written
as

\eqi \left\{
\begin{array}{lll}\lb{gravity}
g_{\rm eq}&=&\rp{GM}{R_{\rm eq}^2}-\omega^2 R_{\rm eq}+\rp{GM}{c^2
R_{\rm eq}^3}\left(4GM-\omega^2 R_{\rm eq}^3\right)
-\rp{2GJ\omega}{c^2 R^2_{\rm eq}},\\\\
g_{\rm pol}&=&\rp{GM}{R_{\rm pol}^2}+\rp{4(GM)^2}{c^2 R_{\rm
pol}^3},
\end{array}
\right. \eqf where the difference between the Earth's radius at
the poles and at the equator is related to the Earth's flattening
$f$ by \eqi\rp{R_{\rm eq}-R_{\rm pol}}{R_{\rm eq}}\equiv f.\eqf
Then, we can write for $\Delta g\equiv g_{\rm pol}-g_{\rm
eq}$\eqi\Delta g\simeq\rp{2GM}{R_{\rm
eq}^2}\rp{f}{(1-f)^2}+\rp{12(GM)^2}{c^2 R_{\rm
eq}^3}\rp{f}{(1-f)^3}+\omega^2\left(R_{\rm
eq}+\rp{GM}{c^2}\right)+\rp{2GJ\omega}{c^2 R^2_{\rm
eq}}.\lb{digi}\eqf
{\tiny\begin{table}\caption{Relevant geodetic
parameters.}\label{para}

\begin{tabular}{lll}
\noalign{\hrule height 1.5pt}\\

Gravitational constant $G$ (kg$^{-1}$ m$^3$
s$^{-2}$) \ct{mohrtay99} &  6.673$\times 10^{-11}\pm 1\times 10^{-13}$\\
Speed of light $c$ (m s$^{-1}$) \ct{mohrtay99} & 2.99792458$\times 10^8$\\
Earth's $GM$ (m$^3$ s$^{-2}$) \ct{groten} & 3.986004418$\times 10^{14}\pm 8\times 10^{5}$ \\
Earth's mean equatorial radius $R_{\rm eq}$ (m) \ct{groten}& $6.3781366\times 10^6\pm 1\times 10^{-1}$ \\
Earth's angular momentum $J$ (kg m$^2$ s$^{-1}$) \ct{iers}& $5.85386532242\times 10^{33}$\\
Earth's flattening $f$ \ct{groten} & $3.352\times 10^{-3}\pm
1.1\times 10^{-10}$\\
Earth's angular speed $\omega$ (rad s$^{-1}$) \ct{groten} &
$7.292115\times 10^{-5}\pm 1\times 10^{-12}$
\\\\

\noalign{\hrule height 1.5pt}
\end{tabular}

\end{table}}

Now we will evaluate the magnitude of the various terms entering
\rfr{digi} in order to compare them with the present-day available
sensitivity of SG. Then, we will also evaluate the systematic
errors induced by the classical terms of \rfr{digi} due to the
uncertainties in the various geodetic parameters of the Earth. Of
course, there are also various classical time-dependent competing
surface gravity effects\footnote{They are, e.g,  the ocean noise,
seismic and normal modes, slow and silent earthquakes, secular
deformations. } spanning a wide range of periodicities from 1 s to
more than 1 year and magnitudes up to $1-10\
\mu$gal=10$^{-5}-10^{-4}$ m s$^{-2}$ which would act as systematic
bias\footnote{It must be noted that the investigated measurement
would be in continuous regime. }. They should be accounted for in
a detailed error budget, which is, however, beyond the scope of
the present paper.
\subsection{The Lense-Thirring component}
By using the values of Table \ref{para} it turns out that the
Lense-Thirring component $\Delta g_{\rm LT}\equiv 2GJ\omega/c^2
R_{\rm eq }^2$ of \rfr{digi} amounts to $1.5\times 10^{-11}$ m
s$^{-2}$, i.e. $\sim$ 1 ngal.
\subsubsection{The systematic errors}
In order to obtain $\Delta g_{\rm LT}$ we should subtract the
first three terms of \rfr{digi} from the measured $\Delta g$. This
can only be done if the uncertainty of the terms to be subtracted
is smaller that the predicted value of the Lense-Thirring
component.

The residual Schwarzschild component would not pose problems.
Indeed, its nominal value is \eqi\Delta g_{\rm Schwarzschild}=
\rp{12(GM)^2}{c^2 R_{\rm
eq}^3}\rp{f}{(1-f)^3}+\rp{GM}{c^2}\omega^2 =3.00\times 10^{-10}\
{\rm m\ s}^{-2}.\lb{dschwa}\eqf According to the uncertainties
released in Table \ref{para} it turns out that $\delta(\Delta g_{
\rm Schwarzschild})\ll \Delta g_{ \rm LT}$. The problems come from
the classical terms. Indeed, \eqi\Delta g_{\rm
Newton}=\rp{2GM}{R_{\rm eq}^2}\rp{f}{(1-f)^2}=6.6130308646\times
10^{-2}\ {\rm m\ s}^{-2}, \eqf which is nine orders of magnitude
larger than the gravitomagnetic term. The errors induced by the
uncertainty in $f$ and the Earth's mean equatorial radius $R_{\rm
eq}$ are the largest ones and amount to \eqi\delta(\Delta g_{\rm
Newton} )|_f=\rp{2GM}{R^2_{\rm eq}}\rp{(1+f)}{(1-f)^3}\delta
f=2.184\times 10^{-9}\ {\rm m}\ {\rm s}^{-2}\eqf and
\eqi\delta(\Delta g_{\rm Newton} )|_{R_{\rm eq}}=\rp{4GMf}{
R^3_{\rm eq}(1-f)^2 }\delta R_{\rm eq}=2.073\times 10^{-9}\ {\rm
m} \ {\rm s}^{-2},\eqf respectively. The centrifugal component
amounts to \eqi\Delta g_{\rm centrifugal}=\omega^2R_{\rm
eq}=3.3906727993\times 10^{-2}\ {\rm m\ s}^{-2}, \eqf with an
error of $\delta(\Delta g_{\rm centrifugal})=1.48\times 10^{-9}$ m
s$^{-2}$. The total error in the classical part of $\Delta g$ is,
thus \eqi\delta(\Delta g)^{\rm class}_{\rm total}\leq 5.737\times
10^{-9}\ {\rm m\ s}^{-2};\lb{errdgtot}\eqf it is two orders of
magnitude larger than $\Delta g_{\rm LT}$. This rules out the
possibility of measuring the gravitomagnetic Lense-Thirring
component of the acceleration of gravity with the present-day SG.
\subsection{The Schwarzschild component}
From \rfr{gravity} it can be noted that an absolute measurement at
the South Pole could allow to detect the gravitoelectric
Schwarzschild component of the acceleration of gravity. Indeed, it
turns out that \eqi g_{\rm pol}^{\rm
Schwarzschild}=\rp{4(GM)^2}{c^2 R^3_{\rm eq}(1-f)^3}=2.752\times
10^{-8}\ {\rm m\ s}^{-2}.\eqf
\subsubsection{The systematic errors}
On the other hand, the errors in the classical part are $\delta
g_{\rm pol}^{\rm class}|_{GM}=1.979\times 10^{-8}$ m s$^{-2}$,
$\delta g_{\rm pol}^{\rm class}|_{R_{\rm eq}}=3.0724\times
10^{-7}$ m s$^{-2}$, $\delta g_{\rm pol}^{\rm class}|_f=1.97\times
10^{-9}$ m s$^{-2}$. The total systematic error would then be
\eqi\delta g^{\rm class}_{\rm pol}|_{\rm total}=3.29\times
10^{-7}\ {\rm m\ s}^{-2},\lb{errgpoltot}\eqf i.e. one order of
magnitude larger than $g_{\rm pol}^{\rm Schwarzschild}$. Note that
\rfr{dschwa} and \rfr{errdgtot} rule also out the possibility of a
measurement of the gravitoelectric component of the acceleration
of gravity between the pole and the equator by one order of
magnitude.
\section{Discussion and conclusions}
In this paper we have preliminarily investigated the feasibility
of an experiment aimed at the measurement of the general
relativistic gravitoelectric and gravitomagnetic components of the
terrestrial acceleration of gravity at the equator and at the
pole. It turns out that they are quite small; the difference
between the gravitoelectric accelerations in the two places
amounts to $3\times 10^{-10}$ m s$^{-2}$ and the difference of the
gravitomagnetic accelerations is $1.5\times 10^{-11}$ m s$^{-2}$,
i.e. 10 ngal and 1 ngal, respectively. Such absolute measurements
could be done, in principle, with SG. However, it must be noted
that the greatest difficulty in the use of the SG as an absolute
instrument is connected with its calibration. Another non-trivial
difficulty is that the effects to be measured are in continuous.
Moreover, the present-day sensitivity of SG is just of the order
of ngal, but only for relative measurements. In regard to
systematic bias, the current uncertainties in the Earth's geodetic
parameters, which enter the classical Newtonian terms to be
subtracted, induce errors 1-2 orders of magnitude larger than the
relativistic ones.
\section*{Acknowledgements}
I thank the anonymous referees for their helpful and important
comments which greatly improved the present paper.
%


\begin{thebibliography}{xxxxx}

\bibitem{soffel}
Soffel, M.H., 1989, {\it Relativity in Astrometry, Celestial
Mechanics and Geodesy}, (Springer, Berlin).

\bibitem{will}
Will, C.M., 2001, {\it Living Rev. Relativ.} {\bf 4}, 4
{\texttt{http://www.livingreviews.org/Articles/Volume4/2001-4will}}

\bibitem{bra}
V.B. Braginsky, V.B., Polnarev, A.G., and Thorne, K.S., 1984, {\it
Phys. Rev. Lett.}, {\bf 53}, 863.

\bibitem{pippard}
Pippard, A.B., 1988, {\it Proc. R. Soc. London A}, {\bf 420}, 81.

\bibitem{bra2}
Braginsky, V.B., Caves, C., and Thorne, K.S., 1977, {\it Phys.
Rev. D}, {\bf 15}, 2047.

\bibitem{vitale}
Cerdonio, M., Prodi, G.A., and Vitale, S., 1988, {\it Gen. Rel.
Grav.}, {\bf 20}, 83.

\bibitem{tartrug02}
Tartaglia, A., and Ruggiero, M.L. 2002, {\it Gen. Rel. Grav.},
{\bf 34}, 1371.

\bibitem{stedman}
Stedman, G.E., Schreiber, K.U., and Bilger, H.R., 2003, {\it
Class. Quantum Grav.}, {\bf 20}, 2527.

\bibitem{ioriopolo}
Iorio, L., 2003, {\it Class. Quantum Grav.}, {\bf 20}, L5.

\bibitem{goodk}
Goodkind, J.M., 1999, {\it Review of Scientific Instruments}, {\bf
70}, 4131.

\bibitem{nanogal}
Francis, O., and van Dam, T., 2002, {\it Metrologia}, {\bf 39},
485.

\bibitem{WIL}
Will, C.M., 1993, {\it Theory and experiment in gravitational
physics}, revised edition, (Cambridge University Press,
Cambridge).


\bibitem{Schw}
Schwarzschild, K., 1916,
{\it Sitzungsberichte der K\"{o}niglich Preussischen Akademie der
Wissenschaften zu Berlin, Phys.-Math. Klasse}, 189 translated and
discussed by Antoci, S., 2003, {\it Gen. Rel. Grav.}, {\bf 35},
951.

\bibitem{leti}
Lense, J., and Thirring, H., 1918,
{\it Phys. Z.} {\bf 19}, 156 translated and discussed by Mashhoon,
B., Hehl, F. W., and Theiss, D. S., 1984,
 {\it Gen. Rel. Grav.}, {\bf 16}, 711.


\bibitem{mohrtay99}
Mohr, P.J., and Taylor, B.N., 1999, {\it J. Phys. Chem. Ref.
Data}, {\bf 28}, 6, p. 1713.

\bibitem{groten}
Groten, E., 1999, {\it Report of the IAG Special Commission SC3,
Fundamental Constants}, XXII IAG General Assembly.

\bibitem{iers}
McCarthy, D. D., and Petit, G., 2004, {\it IERS conventions
(2003)} IERS Technical Note 23 (Verlag des Bundesamtes f\"{u}r
Kartographie und Geod\"{a}sie, Frankfurt am Main).

\end{thebibliography}
\end{document}